\documentclass[showpacs,twocolumn,prl,aps,10pt]{revtex4-1}

\usepackage{graphicx}
\usepackage{amssymb}
\usepackage{amsmath}
\usepackage{color}
\usepackage{soul}
\usepackage{psfrag}
\usepackage{epsfig}
\usepackage{bm}
\usepackage{ifthen}
\usepackage{calc}
\usepackage[latin1]{inputenc}
\usepackage[T1]{fontenc}
\usepackage{amsfonts}
\usepackage{amstext}
\usepackage{pst-all}
\usepackage[english]{babel}
\usepackage{subfigure}
\usepackage{rotating}
\usepackage{textcomp}
\usepackage{hyperref}
\definecolor{darkblue}{rgb}{0.0,0.0,0.3}
\hypersetup{colorlinks,breaklinks,linkcolor=darkblue,urlcolor=darkblue,anchorcolor=darkblue,citecolor=darkblue}

\newcommand{\ket}[1]{| #1 \rangle}

\newcommand{\beq}{\begin{eqnarray}}
\newcommand{\eeq}{\end{eqnarray}}

\newcommand{\eq}[1]{Eq.~(\ref{#1})}

\newcommand{\no}{\nonumber}

\begin{document}

\title{Waiting Time Distributions for the Transport  through  a Quantum Dot Tunnel Coupled to One  normal  and One superconducting lead}
\author{Leila Rajabi}
\affiliation{
School of Chemical and Physical Sciences and MacDiarmid Institute for Advanced Materials and Nanotechnology,
Victoria University of Wellington, P.O. Box 600, Wellington 6140, New Zealand}

\author{Christina P\"oltl}
\email[E-mail: ]{ christina.poeltl@vuw.ac.nz}
\affiliation{
School of Chemical and Physical Sciences and MacDiarmid Institute for Advanced Materials and Nanotechnology,
Victoria University of Wellington, P.O. Box 600, Wellington 6140, New Zealand}

\author{Michele Governale}
\email[E-mail: ]{ Michele.Governale@vuw.ac.nz}
\affiliation{
School of Chemical and Physical Sciences and MacDiarmid Institute for Advanced Materials and Nanotechnology,
Victoria University of Wellington, P.O. Box 600, Wellington 6140, New Zealand}

\date{\today}

\begin{abstract}
We have studied the waiting time distributions (WTDs) for subgap transport through a single-level quantum dot tunnel coupled to one normal and one superconducting lead. The WTDs reveal the internal dynamics of the system, in particular, the coherent transfer of Cooper pairs between the dot and the superconductor. The WTDs exhibit oscillations that can be directly associated to the coherent oscillation between the empty and doubly occupied dot. The oscillation frequency is equal to the energy splitting between the Andreev bound states. These effects are more pronounced when the empty state and double-occupied state  are in resonance. 
\end{abstract}
\maketitle



\begin{figure}[t!]
\includegraphics[width=0.90\columnwidth]{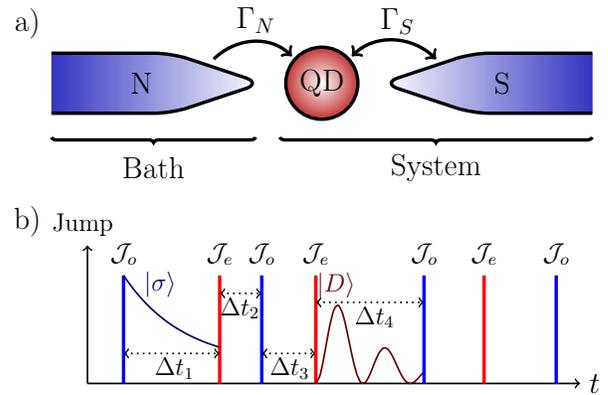}
\caption	{(a) Schematic setup:  quantum dot  (QD) tunnel coupled to one normal lead (N) with tunnel-coupling strength $\Gamma_N$ and one superconducting lead (S) with tunnel-coupling strength $\Gamma_S$. 
Superconducting lead and QD are considered as a combined hybrid {\it system}.  
The normal lead is model as  an electron {\it bath}. 
(b) Time line with jumps processes, a jump $\mathcal J_{\text{o}}$ to the single-occupation sector always has to be followed by a jump $\mathcal J_{\text{e}}$ to the even-occupation sector.  The general waiting time $w(\tau)$ is the distribution of all $\Delta t_m$, $m\in \mathbb N$. In the exemplary sequence of jumps shown, 
the waiting time matrix entries $w_{\text{e},\text{o} }(t)$ and $w_{\text{o}, \text{e} }(t)$ are the distribution of all $\Delta t_{2m-1}$ and $\Delta t_{2m}$. The two curves in time intervals $\Delta t_1$ and $\Delta t_4$, are sketches of the WTDs. 
}
\label{fig:sketch}
\end{figure}

{\it Introduction}--- Electron transport in mesoscopic conductors is an inherently stochastic process \cite{Bl-Bu-noise-rep} and in order to fully characterize it, it is necessary to study the statistics of the transport events \cite{nazarov-book}. A well established theoretical tool is the full counting statistics (FCS) \cite{LeLe93-FCS,LLL96-FCS,BaNa03-FCS, PJS03-FCS,BKR06-GM-FCS,ScBr-BMS-app, Be-Na-FCS, EMA07-Fin-Fre-cou, FNB08-Fin-Fre-cou, MEB10-FCS, MEB11-FCS-non-Mar}. 
The FCS is usually defined in the long-time limit, when the time interval during which transport events are counted is long enough that many particles have passed through the system. The long-time FCS describes fully \textit{zero-frequency} transport quantities, such as the average current, the zero-frequency noise, and higher-order current cumulants. Only recently, the FCS has been extended to the finite frequency domain 
\cite{EMA07-Fin-Fre-cou, FNB08-Fin-Fre-cou, MEB10-FCS, MEB11-FCS-non-Mar}.
Another tool that has recently been employed to characterize mesoscopic transport is the distribution of delay times between subsequent transport events, the 
waiting time distribution (WTD)  \cite{Bra08-wait, WEH08-wait, WMY09-wait, AFB11-wait, AHF12-wait, ThFl12-wait}. 
When memory effects can be neglected, the WTDs can be used as a theoretical tool to evaluate the zero-frequency FCS \cite{Bra08-wait, AFB11-wait}. 
However, their utility lay in the fact that they are particularly suited to study the short-time behaviour of the system.  
This is particularly important for systems that have internal dynamics. For this reason, WTDs have been employed to study double-quantum dots \cite{Bra08-wait, WEH08-wait, WMY09-wait, ThFl12-wait}, where coherent oscillations between states localized in the different dots can occur. 
Quantum dots contacted with superconducting leads \cite{defranceschi10,martin-rodero11}  can show an interesting dynamics due to the coherent exchange of Cooper pairs between the dot and the superconductors. In this particular case, the coherent oscillations occur between dot states with different particle numbers. The microscopic mechanism underlying this effect is  Andreev reflection \cite{degennes63,andreev64}, which leads to the appearance of subgap resonances in the density of states,  so called Andreev bound states, which have been measured by means of transport spectroscopy \cite{pillet10,dirks11}. 
In a quantum dot with Coulomb repulsion that is  tunnel coupled strongly to a superconducting lead and weakly to a normal lead,  a finite pair amplitude can be induced in the dot, which is facilitated by the nonequilibrium due to finite applied voltages \cite{PGK07-SC-GMa,governale08,BGP11-N-QD-S}. 
Such a pair amplitude, describes the coherent exchange of Cooper pairs between the dot and the superconductor and should also be visible in the WTDs.
In the present Letter, we calculate the WTDs for a single-level quantum dot tunnel coupled to one normal and one superconducting lead in the unidirectional transport regime, see Figure \ref{fig:sketch}(a). We find that the WTDs show oscillations which are a signature of the coherent transfer of Cooper pairs back and forth between dot and superconductor. Finally, we discuss possible ways to measure the WTDs without destroying the coherent oscillations in the quantum dot. 

{\it Model}---  
Since we are interested in the subgap Andreev transport between the QD and the superconductor rather than in quasiparticle transport, we assume the superconducting gap to be large.  In this limit, the dot in proximity to the superconductor is described by the effective Hamiltonian  \cite{RoAr00-N-QD-S-effH,BGP11-N-QD-S}
\begin{align}
 H_{\text{eff}}=\varepsilon\sum_{\sigma} d_{\sigma}^{\dagger}{d_\sigma}+U{n_{\uparrow}}{n_{\downarrow}}-\frac{\Gamma_S}{2}(d_{\uparrow}^{\dagger}d_{\downarrow}^{\dagger}+d_{\downarrow}d_{\uparrow}),
\label{eq:effective Hamiltonian-m}
\end{align}
with $ d_{\sigma}^{\dagger}$(${d_\sigma}$) the creation (annihilation) operator of an electron with spin $\sigma=\uparrow , \downarrow$ in the dot and $n_{\sigma}$ the corresponding number operator.  The single particle energy $\varepsilon$ is assumed to be independent of the spin and $U$ denotes  the on site Coulomb repulsion.   
The last term in the Hamiltonian Eq.~(\ref{eq:effective Hamiltonian-m}) is due to the coupling to the superconducting lead and accounts for the coherent tunneling of Cooper pairs in and out of the dot. The prefactor  $\Gamma_S$ is the coupling strength between the dot and the superconducting lead. 
The eigenstates of $H_{\text{eff}}$ with an odd number of electrons are the singly-occupied dot states $\ket{\sigma}=d_\sigma^\dagger \ket{0}$  with energies $\varepsilon$, where $\ket{0}$ is the empty-dot state. Those with an even number of electrons  are the Andreev-bound states 
$\ket{\pm} ={\frac1{\sqrt{2}}}{\sqrt{1\mp\frac{\delta}{2\varepsilon_A}}\ket{0}}\mp{\frac1{\sqrt{2}}}{\sqrt{1\pm\frac{\delta}{2\varepsilon_A}}\ket{D}}$, with eigenenergies $\varepsilon_{\pm}={{\delta}/2}{\pm} {\varepsilon_A}$, where $\ket{D}=d_{\uparrow}^\dagger d_{\downarrow}^\dagger\ket{0}$. Here, ${\delta}={2\varepsilon}+U$ is the detuning between empty and double-occupied state, and $2{\varepsilon_A}=\sqrt{{{\delta}^2}+{\Gamma_S^2}}$ is the splitting between $\ket{+}$ and $\ket{-}$. The normal lead is described by $H_{N}=\sum_{k{\sigma}}\varepsilon_{{N}k} c_{{N}k{\sigma}}^{\dagger}c_{{N}k{\sigma}}$, where $c_{{N}k{\sigma}}^{\dagger}$($c_{N k{\sigma}}$) is the creation (annihilation) operator of an electron with quantum number $k$ and  energy  $\varepsilon_{N k}$. 
The electron distribution in the normal lead is the Fermi function $f(\omega)=(e^{\beta (\omega -\mu)}+1)^{-1}$ where $\mu$ is the chemical potential and $\beta= (k_B T)^{-1}$, with $T$ being the temperature and $k_B$ the Boltzman's constant. 
The proximised  dot is coupled to the normal lead  by means of the tunneling Hamiltonian $H_T=\sum_{k\sigma} V_{N} c_{Nk\sigma}^{\dagger}d_{\sigma}+\text{H.c.}$, where $V_N$ is the tunneling amplitude. 
We define the tunneling coupling strength $\Gamma_N=2\pi \nu_{\text{N}} |V_{\text{N}}|^2$, where $\nu_{\text{N}}$ is the density of states in the normal lead, which is assumed to be constant.   
We trace out the reservoir degrees of freedom of the normal lead to obtain a generalized master equation \cite{Breuer,BKR06-GM-FCS,governale08} for the reduced density matrix of the system, $\rho$.  
Since the waiting times are defined for unidirectional transport, we set the chemical potential of the normal lead to be much larger than all relevant energy scale of the system (apart from the superconducting gap). 
In first order, in the coupling strength $\Gamma_{\text{N}}$, the generalised master equation is Markovian and in Liouville space, it can be written 
\cite{BaNa03-FCS,EMA07-Fin-Fre-cou}  as $\dot{\rho}=\mathcal{W} \rho$, where $\mathcal{W}$ is the Liouvillian or kernel.


Introducing  a counting field $\chi$ \cite{LeLe93-FCS,LLL96-FCS,BaNa03-FCS,BKR06-GM-FCS,MEB10-FCS} which counts  
the electrons tunneling  out of the normal lead into the proximised dot,  we can 
write the counting-field dependent equations of motion as $\dot{\rho}(\chi)=(\mathcal{W}_0+e^{-i \chi}\mathcal{J}) \rho(\chi)$, where $\mathcal{J}$ is the jump operator describing the particles tunneling into 
the system from 
the normal lead. The kernel $\mathcal{W}_{0}$ describes the dynamics of the system in interaction with the bath without the jumps and contains both terms in zeroth and first order in $\Gamma_{\text{N}}$. 
The explicit expression for the kernel $\mathcal{W}_0$ and the jump operator $\mathcal{J}$ can be found in the Supplementary information. 

\begin{figure}[t!]
 \centering
\includegraphics[width=0.95\columnwidth]{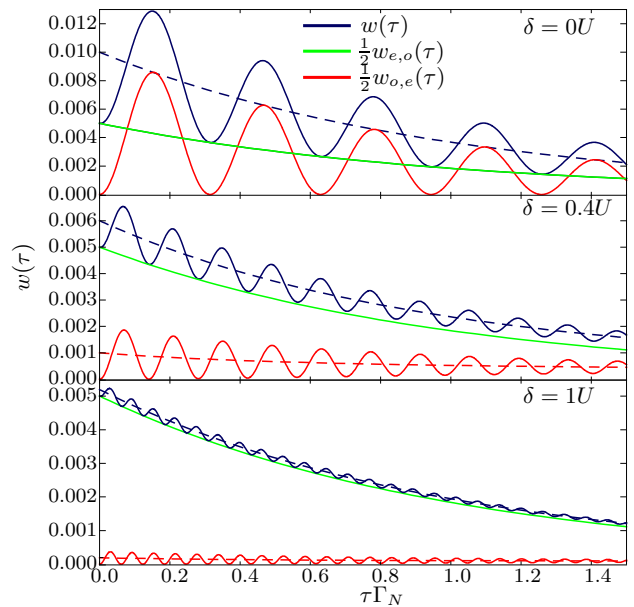}
\caption	{The waiting time distributions $w_{\text{e},\text{o}}(\tau)$, $w_{\text{o} ,\text{e}}(\tau)$, and $w(\tau)$ for $\delta=0U,\ 0.4U,\ 1U $. For ease of visualization, $w_{\text{e},\text{o}}(\tau)$, $w_{\text{o} ,\text{e}}(\tau)$ have been multiplied by a factor $1/2$.  
For comparison, the dashed curves for $w_{\text{o} ,\text{e}}(\tau)$  and $w(\tau)$ are computed without the inclusion of the off diagonal elements of the reduced density matrix between the Andreev bound states. 
Parameters: $\Gamma_N=0.01U$ and $\Gamma_S=0.2U$.
}
\label{fig:wait}
\end{figure}

\textit{Results}--- The waiting time distribution, that is  the probability distribution, for two subsequent jumps to be separated by the time interval $\tau$ when the system is initially in the stationary state described by  $\rho_S$, is given by the expression\cite{Bra08-wait, WEH08-wait, WMY09-wait, ThFl12-wait}:
\begin{align}
\label{wtd-def}
w(\tau)=\frac{ \text{Tr}[\mathcal J e^{\mathcal W_0 \tau} \mathcal J \rho_S  ]}{\text{Tr}[\mathcal J \rho_S  ]}. 
\end{align}

The system under consideration is a multireset system, since it can host more than a single excess electron. For a multireset system, in order to be able to recover the long-time FCS, one identifies different types of jumps and defines a  matrix $W(\tau)$ \cite{Bra08-wait} with elements $w_{k,l}$:
\begin{align}
w_{k,l}(\tau)=\frac{ \text{Tr}[\mathcal J_k e^{\mathcal W_0 \tau} \mathcal J_l \rho_S  ]}{\text{Tr}[\mathcal J_l \rho_S  ]},
\end{align}
and $\mathcal{J}=\sum_k \mathcal{J}_{k}$. The entries of the waiting time matrix are WTDs for a jump of the type $\mathcal{J}_k$ given that the last jump was of the type $\mathcal{J}_l$.
For the hybrid system under consideration we define two different jumps: $\mathcal{J}_{\text{o}}$ which describes jumps from the  Andreev bound states $\ket{\pm}$ (even-occupation sector) to  the single-occupied states $\ket{\sigma}$ (odd-occupation sector) and  $\mathcal{J}_{\text{e}}$ describing the opposites process.  An example of sequence of jumps for our system is shown in panel b) of Fig.~\ref{fig:sketch}; notice that the jumps $\mathcal{J}_{\text{o}}$ and $\mathcal{J}_{\text{e}}$ alternate regularly. 
The waiting time matrix in our case has the dimension 2.  Instead of working in the time domain, we find more convenient to express our results in Laplace space; 
the Laplace transform  of an arbitrary function $g(\tau)$ is defined as $\hat{g}(z)= \int^\infty_0 d\tau e^{-z \tau} g(\tau)$.   
Since in our system, subsequent jumps of the same type are impossible, we find $\hat{w}_{\text{e}, \text{e}}(z)=\hat{w}_{\text{o}, \text{o}}(z)=0$.  
The off diagonal elements are given by    
\begin{subequations}
\label{result-laplace}
\begin{align}
\hat{w}_{\text{o}, \text{e}}(z) & = \frac{\Gamma _N \Gamma _S^2 \left(\Gamma _N+z\right)}{\Gamma _N^2 \Gamma _S^2+z^2 
\alpha_2+2 z \Gamma _N \alpha_1 +4 z^3 \Gamma _N+z^4}, \\
\hat{w}_{\text{e}, \text{o}}(z)& = \frac{\Gamma _N}{\Gamma _N+z}, 
\end{align}
\end{subequations}
with $\alpha_1=\delta ^2+\Gamma _N^2+\Gamma _S^2 $, $\alpha_2=\delta ^2+5 \Gamma _N^2+\Gamma _S^2 $. Here, it is important to emphasize that in order to describe correctly the short-time dynamics of the system, it is necessary to include the off diagonal elements of the reduced density matrix between the Andreev bound states, which we refer to as the {\it coherences}. 
Since in the hybrid system a jump to the single-particle sector $\ket{\sigma}$ has to be followed by a jump to the $\ket{\pm}$ sector and vice versa, the general waiting time  of Eq.~(\ref{wtd-def}) reads $w(\tau)=\frac{1}{2}\left({w}_{\text{e}, \text{o}}(\tau) +{w}_{\text{o} ,\text{e}}(\tau)\right)$.

The waiting time  ${w}_{\text{e}, \text{o}}(\tau)$ is given by the expression ${w}_{\text{e}, \text{o}}(\tau)=\Gamma_N \exp{\left(-\Gamma_N \tau\right)} $. This can be easily understood since after a jump $\mathcal{J}_{\text{o}}$, the dot is in a singly-occupied state and, therefore,  decoupled from the superconductor. The next jump will then be a Poissonian process with rate $\Gamma_N$. 
The situation is different for  ${w}_{\text{o},\text{e}}(\tau)$. The jump $\mathcal{J}_{\text{e}}$ will bring the system from $\ket{\sigma}$ to $\ket{D}$ which  is a coherent superposition of the two eigenstates  
$\ket{+}$ and $\ket{-}$. Therefore we expect coherent oscillations with a frequency that in zeroth order in $\Gamma_N$ is given by the Andreev-bound state splitting $2\varepsilon_{\text{A}}$.  
Although we have the full result in Eq.~(\ref{result-laplace}), it is still instructive to look at the two limiting cases when the proximity effect is in resonance, $\delta=0$, and off resonance,  $\delta\gg \Gamma_S$.  On resonance, $2 \varepsilon_\text{A}=\Gamma_S$ and the waiting time  ${w}_{\text{s},\text{A}}(\tau)$ reduces to 
 \begin{align}
 \nonumber
  {w}_{\text{o} ,\text{e}}(\tau)\Big|_{\delta=0 }&= \frac{\Gamma _N \Gamma _S^2}{\Gamma
   _S^2-\Gamma _N^2} e^{- \Gamma _N\tau} \left[1-\cos \left( \sqrt{\Gamma _S^2-\Gamma _N^2}\tau\right)\right]\\ 
 &\approx \Gamma_N e^{- \Gamma _N\tau} \left[1-\cos (\Gamma _S \tau)\right],
  \end{align}
where in the second equality we have made use of the fact that $\Gamma_N\ll\Gamma_S$. 
On the other hand, off resonance, $2 \varepsilon_\text{A}\approx|\delta| $, we find for short times $\tau \Gamma_N \ll 1	 $
\begin{subequations}
 \begin{align}
 {w}_{\text{o},\text{e}}(\tau)&\approx  \frac{\Gamma _N \Gamma _S^2 }{\delta^2} e^{-\tau  \Gamma _N} \left[1- \cos (|\delta|  \tau) \right]. \label{off-short}
\intertext{and for long times $\tau \Gamma_N \gg 1$ }
  {w}_{\text{o},\text{e}}(\tau) &\approx \frac{\Gamma _N \Gamma _S^2}{2 \delta^2} 
 e^{-\tau \frac{\Gamma _N \Gamma _S^2}{2 \delta^2} }. \label{off-long}
\end{align}
\end{subequations}
The off resonance  long time behaviour is consistent with the picture of Poissonian tunneling of Cooper pairs found in Ref.~\cite{BGP11-N-QD-S}. 
The full result for the  waiting times $w(\tau)$,  ${w}_{\text{e}, \text{o}}(\tau)$ and ${w}_{\text{o}, \text{e}}(\tau)$ obtained by the inverse Laplace transform of Eq.~(\ref{result-laplace}) is shown in Figure \ref{fig:wait}  for different detunings $\delta$, including intermediate ones. The WTD ${w}_{\text{o},\text{e}}(\tau)$ always start from zero for $\tau=0$, since after the jump $\mathcal{J}_{\text{e}}$ the system is in $\ket{D}$ and any further transport is blocked until the two electrons in the dot are transferred coherently to the superconductor.  We notice that, as the detuning increases, the amplitude of the oscillations is suppressed and the frequency of the oscillations is increased. 
From the discussion above, it is clear that in order to describe the oscillations, it is necessary to include the coherences between the Andreev-bound states. Not including the coherences gives only the exponential behaviour and does not resolve the oscillating terms. 
This is demonstrated in  Figure \ref{fig:wait}, where  we also show $w(\tau)$ and ${w}_{\text{o},\text{e}}(\tau)$ obtained without the inclusion of the coherences. 
\begin{figure}[t!]
 \centering
\includegraphics[width=0.95\columnwidth]{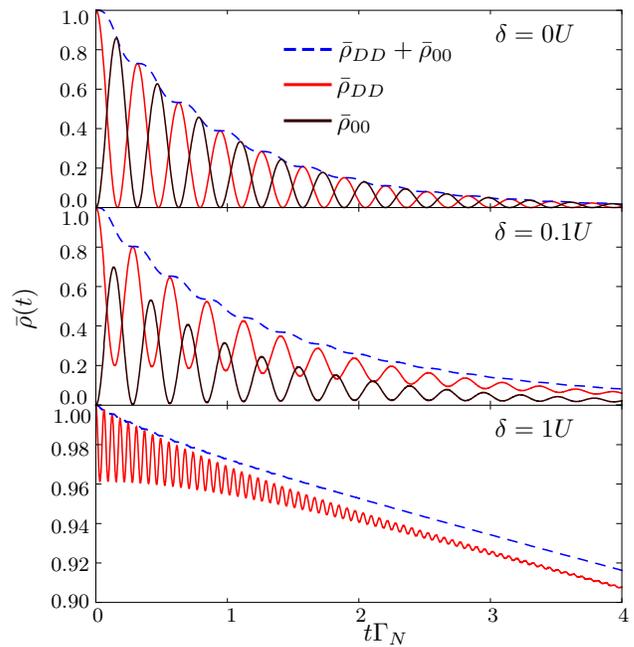}
\caption	{Time evolution of  $\bar{\rho}_{DD}+\bar{\rho}_{00}$, $\bar{\rho}_{DD}$  and $\bar{\rho}_{00}$ obtained by the time evolution with the Liouvillian without jumps $\mathcal{W}_0$ when the system is initially prepared in the doubly occupied state for $\delta=0U$, $0.1U$, $1U$.  
Parameters: $\Gamma_N=0.01U$ and $\Gamma_S=0.2U$.         }
\label{fig:bol}
\end{figure}

In order to understand the features of $w_{\text{o}, \text{e}}(\tau)$, we investigated the time evolution of $\bar{\rho}(t)=e^{\mathcal{W}_0 t}\rho_D$,  that is the evolution of the system without jumps when at $t=0$,  the system is in the doubly occupied state. The quantity $\bar{\rho}(t)$ are the elements of the density matrix under the condition that no jumps have occurred in the time interval t.  
We use the localised basis for the dot and write $\bar{\rho}=\{\bar{\rho}_{00},\bar{\rho}_{\uparrow\uparrow},\bar{\rho}_{\downarrow\downarrow},\bar{\rho}_{DD},\bar{\rho}_{D0},\bar{\rho}_{0D}  \}$, with  $\bar{\rho}_{\uparrow\uparrow}(t)=\bar{\rho}_{\downarrow\downarrow}(t)=0$ since no jump has occurred.  
The time evolution of $\bar{\rho}_{DD}$ and $\bar{\rho}_{00}$ clearly governs the evolution of $w_{\text{o}, \text{e}}(\tau)$. 
Figure \ref{fig:bol} shows the time evolution of $\bar{\rho}_{DD}+\bar{\rho}_{00}$, $\bar{\rho}_{DD}$ and $\bar{\rho}_{00}$ for different detunings $\delta$. 
As the system is initially in the doubly occupied state, we have as initial values $\bar{\rho}_{DD}(t\mathord=0)+\bar{\rho}_{00}(t\mathord=0)=1$, $\bar{\rho}_{DD}(t\mathord=0)=1$, and $\bar{\rho}_{00}(t\mathord=0)=0$. All the elements of $\bar{\rho}$ are decaying since the probability of having no jumps decays with time and eventually goes to zero as times goes to infinity \cite{note1}. 
On resonance, the local maxima of $\bar{\rho}_{DD}(t)$ are equal to $\bar{\rho}_{DD}(t)+\bar{\rho}_{00}(t)$ at all times, since the coherent dynamics govern the evolution of $\bar{\rho}$ and the  behaviour of the WTDs at all times. 
This is not the case off resonance in the long-time limit ($t\Gamma_N\gg 1$), where the coherences are suppressed and the WTDs show only a Poissonian decay, see \eq{off-long}.

The waiting time matrix in Laplace space  $\hat{W}(z)$ can be used to derive  the long-time FCS. In particular, the cumulant generating function is the solution  
of the equation $\det [ e^{i  \chi } - \hat{W}(z) ] = 0, \label{wait-FCS}$
that fulfils $z_0(\chi\mathord=0)=0$.
We find 
$z_0= -\Gamma _N+\frac{\sqrt{\Gamma _N^2-\delta ^2-\Gamma _S^2+\sqrt{\left(\delta ^2+\Gamma _N^2+\Gamma _S^2\right)^2+4 \left(e^{-2 i \chi
   }-1\right) \Gamma _N^2 \Gamma _S^2}}}{\sqrt{2}} $, 
which  reduces in the limit of small $ \Gamma_N$ to the result of Ref.  \cite{BGP11-N-QD-S} obtained without the inclusion of the coherences  between $\ket{\pm}$,
$
z_0 =\Gamma_N \left( \sqrt{\frac{\delta ^2+e^{-2 i \chi } \Gamma _S^2}{\delta ^2+\Gamma _S^2}}-1\right ).
$

\begin{figure}[t!]
 \centering
{\includegraphics[width=0.95\columnwidth]{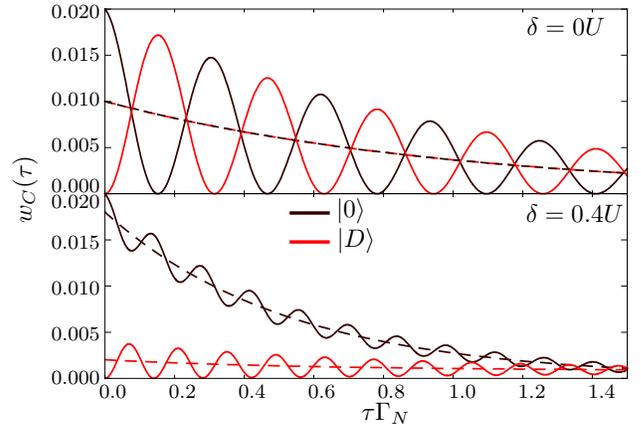}}
\caption	{Conditional waiting times $\hat{w}_{C,\ket{0}}(z)$ (system initially empty $\ket{0}$) and $\hat{w}_{C,\ket{D}}(z)$ (system initially double occupied $\ket{D}$) for $\delta\mathord=0U$ and $\delta\mathord=0.4U$ at $\Gamma_N\mathord=20\Gamma_S$. The dashed curves are calculated without the inclusion of the coherent dynamics. 
        }
\label{fig:wait-C}
\end{figure}

We also define the conditional waiting time as 
\begin{align}
w_{C,\xi}(\tau)=\text{Tr}[\mathcal J e^{\mathcal W_0 \tau} \rho_\xi  ],
\end{align}
which is the probability distribution for the next jump to happen after the time $\tau$ given that the system has been prepared initially in the state $\ket{\xi}$.  
The conditional waiting time can be used to measure the entries of the waiting-time matrix directly. 
Since after a jump from the odd sector to the $\ket{\pm}$ sector the system is in the double-occupied state, we have that the conditional waiting time $\hat{w}_{C,\ket{D}}(z)$ is simply $\hat{w}_{C,\ket{D}}(z)=\hat{w}_{\text{o}, \text{e}}(z)$.  Similarly, we find $\hat{w}_{C,\ket{\sigma}}(z)=\hat{w}_{\text{{e},\text{o}}}(z)$, with the difference that a single-occupied state has no coherent evolution in time, such that only the next jump changes the system state. 
The conditional waiting time distribution for the system initially in the empty state is an independent quantity from the entries of the waiting-time matrix and reads
  \begin{align}
\hat{w}_{C,\ket{0}}(z) & = \frac{\Gamma _N \left(\left(\Gamma _N+z\right) \left(2 z \left(\Gamma _N+z\right)+\Gamma _S^2\right)+2 \delta ^2
   z\right)}{\Gamma _N^2 \Gamma _S^2+z^2 
\alpha_2+2 z \Gamma _N \alpha_1 +4 z^3 \Gamma _N+z^4}. \no
\end{align} 
Figure \ref{fig:wait-C} shows the conditional waiting times for the system being prepared in the empty state ${w}_{C,\ket{0}}(\tau)$  and the double-occupied state ${w}_{C,\ket{D}}(\tau)$. 
 Obviously, ${w}_{C,\ket{D}}(0)=0$ since the system is in the double-occupied state and transport is blocked, while ${w}_{C,\ket{0}}(0)$ is at the maximum value since the system is in the empty state and by construction no electron could have tunneled at an earlier time as $\tau=0$. We also show the curves obtained without inclusion of coherent dynamics.

Finally, we discuss the possibility to measure the WTDs for the system under consideration. The WTDs for unidirectional transport  though a quantum dot can be measured by monitoring the occupation of the dot by means of a charge detector, e.g., a nearby capacitively-coupled  quantum point contact (QPC) \cite{gustavsson-06,fricke-07,fujisawa-06,flindt-09,GLS-09-FCS-Exp-Rep,choi-12,ubbeholde-12,VES-04}. 
In the present system, the situation is slightly more complex as  Cooper pairs oscillate coherently back and forth between the dot and the superconducting lead, and a charge  detector  would decohere  these oscillations.
The ideal detector to measure the WTDs in this system  is a device that measures whether an unpaired spin is present on the dot. As a model, let us assume that the current in the detector is of the form $I_{det}=i_s (n_{\uparrow}+ n_{\downarrow}-2n_{\uparrow}n_{\downarrow})$. The detector current is equal to $i_s$ if the dot is in the single-occupied state, and it vanishes if the dot is in the even-occupation sector. The detector current as a function of time corresponds to a time trace for the jumps $\mathcal{J}_{\text{e}}$ and $\mathcal{J}_{\text{o}}$. The distribution of the length of the time intervals with current $i_s$ is the WTD $w_{\text{e},\text{o}}(\tau)$; similarly, the distribution of the length of the  time intervals with zero current is the WTD $w_{\text{o},\text{e}}(\tau)$. Since such a detector is not able to discriminate between the states $\ket{0}$ and $\ket{D}$, it does not introduce decoherence in the coherent tunneling of Cooper pairs. 
%
In order to observe the WTD oscillations, their frequency,  approximately  $\Gamma_S$ close to resonance, needs to be within the detector bandwidth. 
If one considers a state-of-the art QPC charge detector, this requirement is fulfilled by choosing  $\Gamma_S\mathord\approx 100$ kHz \cite{note-bandwidth}. The coupling to the normal lead should be weaker, for example, $\Gamma_N\mathord\approx 10$kHz;  which is within the limit of current experiments \cite{gustavsson-06,GLS-09-FCS-Exp-Rep}. Temperature will need to be larger than $\Gamma_N$ but otherwise as small as possible to avoid further decoherence (standard cryogenic temperatures in the range of few tens of mK will be adequate). 
  Realistic parameters for a quantum dot in the single-level regime are: single-particle level spacing of order $10$ meV and $U\mathord\approx 1$ meV.


{\it Conclusion}--- 
In this Letter, we have calculated the WTDs of a single-level quantum dot coupled to one normal and one superconducting lead in the regime of strong coupling to the superconductor. The WTDs reveal features which are directly related to the Andreev-bound states. In particular, we found that the WTDs oscillate with a frequency equal to the Andreev bound state splitting. The amplitude of the oscillation is maximal when the proximity effect is on resonance  and probes the pair amplitude in the dot. The time scales associated to the coherent Cooper pair oscillations do not appear in the zero-frequency FCS. The entries of the waiting-time matrix can be measured directly with a device that senses the presence of an unpaired spin in the dot.

{\it Acknowledgements}--- We acknowledge fruitful discussions with T.~Brandes, M.~B\"uttiker, C.~Flindt, J.~K\"onig, and K. Thomas.

\newpage
\setcounter{equation}{0}
\renewcommand{\theequation}{S.\arabic{equation}}
\onecolumngrid
\section*{Supplementary information}

The generalised master equation for a nanostructure tunnel-coupled to a fermonic bath is in general non Markovian \cite{Breuer_S,BKR06-GM-FCS_S}. It becomes Markovian when the coupling to the bath $\Gamma_N=2\pi \sum_k| V_{N }|^2 \delta(\omega-\varepsilon_{Nk})$ is taken into account in first order in perturbation theory or in the high-bias limit ($\mu \rightarrow \pm \infty$) when electronic transport is unidirectional. Therefore, for the regime considered here, the kernel is Markovian and the generalised master equation is of Linblad form and  can be written as
\begin{align}
 \dot{\rho} &= \mathcal W  \rho =-i[H_S,\rho]+\mathcal{D}(\rho),
 \end{align}
 where $H_S$ is the system Hamiltonian,  $\mathcal{D}(\rho)$  the Lindblad {\it dissipator} and we have set $\hbar=1$. 
For the system under consideration, we have $H_S=H_{\text{eff}}$  and in the high bias limit ($\mu\rightarrow\infty$) the dissipator reads 
\begin{align}
\mathcal{D}(\rho)=\sum_\sigma \Gamma_N \left(e^{-i\chi}d^\dagger_\sigma \rho d_\sigma-\frac{1}{2}(\rho d_\sigma d^\dagger_\sigma +d_\sigma d^\dagger_\sigma \rho )\right),
 \end{align}
 where we have introduced the counting field $\chi$ that counts electrons tunneling out of the normal lead. 
The first part of the dissipator describes a measured change in the number of particles of the normal lead, i.e. a jump event. The second part describes dissipation of the system due to the coupling to the normal lead without jumps. 
The kernel is then rewritten as  $\mathcal W(\chi)=\mathcal W_0+e^{-i \chi} \mathcal J$, where $\mathcal J$ is the jump operator. We separate the jump operator $\mathcal{J}=\mathcal J_e +\mathcal{J}_o$, into a part $\mathcal J_e $ describing jumps to the the Andreev-bound states (even) sector and a part $\mathcal{J}_o$ describing jumps to the single-occupied (odd) sector. To make this separation we introduce a different counting field for each type of jump and we rewrite the kernel as 
\begin{align}
\mathcal W(\chi_e,\chi_o)=\mathcal W_0+e^{-i \chi_e} \mathcal J_e+e^{-i \chi_o} \mathcal J_o.  
\end{align}
The kernel takes a simple form in the basis $\{\rho_{00},\rho_{\uparrow\uparrow},\rho_{\downarrow\downarrow},\rho_{DD},\rho_{D0},\rho_{0D} \}$, where $\rho_{\xi' \xi}=\langle\xi'|\rho\ket{\xi}$ and the states $\ket{\xi}$ and $\ket{\xi'}$ belong to the localised dot basis $\left\{\ket{0},\ket{\uparrow},\ket{\downarrow},\ket{D}\right\}$. With this choice of basis in Liouville space, the kernel reads 
 \begin{align}		
 \label{kernel-local}
\mathcal W(\chi_o,\chi_e)	= 
\begin{pmatrix}
  -2 \Gamma _N & 0 & 0 & 0 & \frac{i \Gamma _S}{2} & -\frac{i \Gamma _S}{2} \\
 \Gamma _N e^{-i \chi_o } & -\Gamma _N & 0 & 0 & 0 & 0 \\
 \Gamma _N e^{-i \chi_o } & 0 & -\Gamma _N & 0 & 0 & 0 \\
 0 & \Gamma _N e^{-i \chi_e }& \Gamma _N e^{-i \chi_e }& 0 & -\frac{i \Gamma _S}{2} & \frac{i \Gamma _S}{2} \\
 \frac{i \Gamma _S}{2} & 0 & 0 & -\frac{i \Gamma _S}{2} & -i \delta -\Gamma _N & 0 \\
 -\frac{i \Gamma _S}{2} & 0 & 0 & \frac{i \Gamma _S}{2} & 0 & i \delta -\Gamma _N \\
\end{pmatrix}
.	
 \end{align}
From Eq~(\ref{kernel-local}), $\mathcal W_0$ and the jump operators $\mathcal J_e$ and $\mathcal J_o$ can be read off. 

For the sake of completeness we also give the expression for the kernel  in the basis of the eigenstates of $H_{\text{eff}}$: $\left\{\ket{\uparrow},\ket{\downarrow},\ket{+}\ket{-}\right\}$. Using the basis $\{\rho_{\uparrow\uparrow},\rho_{\downarrow\downarrow},\rho_{--},\rho_{++},\rho_{-+},\rho_{+-} \}$ in Liouville space, we obtain 
\begin{align}		
 \mathcal W(\chi_o,\chi_e)= 
 \begin{pmatrix}
 -\Gamma _N & 0 & \frac{\Gamma _N \left(\delta +2 \varepsilon _A\right)}{4 \varepsilon _A}  e^{-i \chi_o}& -\frac{\Gamma _N \left(\delta -2 \varepsilon _A\right)}{4 \varepsilon _A} e^{-i \chi_o}&
   \frac{\Gamma _N \Gamma _S}{4 \varepsilon _A} e^{-i \chi_o}& \frac{\Gamma _N \Gamma _S}{4 \varepsilon _A} e^{-i \chi_o}\\
 0 & -\Gamma _N & \frac{\Gamma _N \left(\delta +2 \varepsilon _A\right)}{4 \varepsilon _A} e^{-i \chi_o}& -\frac{\Gamma _N \left(\delta -2 \varepsilon _A\right)}{4 \varepsilon _A} e^{-i \chi_o}&
   \frac{\Gamma _N \Gamma _S}{4 \varepsilon _A} e^{-i \chi_o}& \frac{\Gamma _N \Gamma _S}{4 \varepsilon _A} e^{-i \chi_o}\\
 -\frac{\Gamma _N \left(\delta -2 \varepsilon _A\right)}{4 \varepsilon _A} e^{-i \chi_e}& -\frac{\Gamma _N \left(\delta -2 \varepsilon _A\right)}{4 \varepsilon _A} e^{-i \chi_e}& -\frac{\Gamma _N
   \left(\delta +2 \varepsilon _A\right)}{2 \varepsilon _A} & 0 & -\frac{\Gamma _N \Gamma _S}{4 \varepsilon _A} & -\frac{\Gamma _N \Gamma _S}{4 \varepsilon _A} \\
 \frac{\Gamma _N \left(\delta +2 \varepsilon _A\right)}{4 \varepsilon _A}e^{-i \chi_e} & \frac{\Gamma _N \left(\delta +2 \varepsilon _A\right)}{4 \varepsilon _A} e^{-i \chi_e}& 0 & \frac{1}{2}
   \Gamma _N \left(\frac{\delta }{\varepsilon _A}-2\right) & -\frac{\Gamma _N \Gamma _S}{4 \varepsilon _A} & -\frac{\Gamma _N \Gamma _S}{4 \varepsilon _A} \\
 -\frac{\Gamma _N \Gamma _S}{4 \varepsilon _A}e^{-i \chi_e} & -\frac{\Gamma _N \Gamma _S}{4 \varepsilon _A} e^{-i \chi_e}& -\frac{\Gamma _N \Gamma _S}{4 \varepsilon _A} & -\frac{\Gamma _N \Gamma
   _S}{4 \varepsilon _A} & 2 i \varepsilon _A-\Gamma _N & 0 \\
 -\frac{\Gamma _N \Gamma _S}{4 \varepsilon _A} e^{-i \chi_e}& -\frac{\Gamma _N \Gamma _S}{4 \varepsilon _A} e^{-i \chi_e}& -\frac{\Gamma _N \Gamma _S}{4 \varepsilon _A} & -\frac{\Gamma _N \Gamma
   _S}{4 \varepsilon _A} & 0 & -\Gamma _N-2 i \varepsilon _A \\
\end{pmatrix}.
\label{C-kernel}
 \end{align}
This representation of the kernel can also be easily obtained by means of a diagrammatic real-time expansion \cite{futterer}.



\begin{thebibliography}{5}
\bibitem{note1} Technically this is due to the fact that  the dynamics of 
$\bar{\rho}(t)$ is not trace conserving, as it is not the evolution with the full Liouvillian.
\bibitem{note-bandwidth} In Ref.~\cite{ubbeholde-12}, a QPC-charge detector with a bandwidth of 300 kHz is experimentally demonstrated.
\end{thebibliography}
\end{document}